\documentclass[12pt,a4paper]{article}
\usepackage[latin1]{inputenc}
\usepackage{amsfonts}

\usepackage{epsf,array,amsmath,amssymb,endnotes}
\usepackage{graphicx,delarray,stmaryrd}
\usepackage{rotating}

\newtheorem{theorem}{Theorem}

\newtheorem{proposition}{Proposition}

\newcommand{\jx}{ x}
\newcommand{\jy}{ y}
\newcommand{\jz}{z}
\newcommand{\jp}{ p }

\newcommand{\jX}{X}
\newcommand{\jY}{Y}
\newcommand{\jZ}{ Z }
\newcommand{\jxi}{ \xi}
\newcommand{\jmu}{ \mu }
\newcommand{\jve}{\varepsilon }
\newcommand{\jW}{ W }
\newcommand{\ja}{ a }
\newcommand{\jb}{ b }
\newcommand{\jv}{ v }
\newcommand{\jw}{ w }

\newcommand{\figcaption}{\def\@captype{figure}\caption}

\newcommand{\sX}{{\cal X}}

\newcommand{\half}{ {\scriptstyle \frac{1}{2}} \, }
\newcommand{\R}{{\mathbb R}  }

\newcommand{\nnt}{t_n}

\newcommand{\bea}{\begin{eqnarray}}
\newcommand{\eea}{\end{eqnarray}}
\newcommand{\beast}{ \begin{eqnarray*} }
\newcommand{\eeast}{ \end{eqnarray*} }

\newcommand{\be}{\begin{equation}}
\newcommand{\ee}{\end{equation}}

\author{L.~C.~G.  Rogers
\\
University of Cambridge
\\
First version: September 2008}
\title{Least-action filtering}
\begin{document}
\maketitle

\bibliographystyle{plain}

\begin{abstract}
This paper presents an approach to estimating
a hidden process in a continuous-time
setting, where the hidden process  is a diffusion.   
The approach is simply to minimize the negative log-likelihood of
the hidden path, where the likelihood is expressed relative to 
Wiener measure.  This negative log-likelihood is the action integral
of the path, which we minimize by calculus of variations.  We then 
perform an asymptotic maximum-likelihood analysis to understand
better how the actual path is distributed around the least-action
path; it turns out that the actual path can be expressed (approximately)
as the sum of the least-action path and a zero-mean Gaussian 
process which can be specified quite explicitly.   Numerical solution
of the ODEs which arise from the calculus of variations is often
feasible, but is complicated by the shooting nature of the problem,
and the possibility that we have found a local but not global minimum.
We analyze the situations when this happens, and provide  effective
numerical methods for studying this.  We also show how the methodology
works in a situation where the hidden positive diffusion acts as the 
random intensity of a point process which is observed; here too it is 
possible to estimate the hidden process.
\end{abstract}

\section{Introduction.}\label{intro}
The basic problem tackled in this paper is to try to estimate
the hidden part $(\jX_t)_{0 \leq t \leq T}$ of a vector\footnote{
We use the Scilab/Matlab notation, where $\jz=[\jx;\jy]$ denotes 
the column vector $\jz$ formed by stacking the column 
vector $\jx$ above the column vector $\jy$.
}
diffusion  process $\jZ_t \equiv [\jX_t;\jY_t]$ given the observations
$(\jY_t)_{0 \leq t \leq T}$. 
Of course, this is an idealized question, because in practice we
would only observe the signal at discrete instants of time,
which for simplicity we will assume are equally spaced with
a spacing of $h>0$.  It would then in principle be possible to 
write down the likelihood\footnote{
We suppose that $T = Nh$.
} of $(\jZ_{nh})_{ 0 \leq n \leq N}$, and maximize this over
the unobserved variables $(\jX_{nh})_{ 0 \leq n \leq N}$,
with the $\jY$-values known and fixed.

 There are two quite 
substantial obstacles on this path. The first is that (except
in a few special examples) there is no closed-form expression
for the transition density of a general diffusion, so evaluating
the likelihood is already problematic. The second is that we 
are faced with a maximization over a space of dimension 
proportional to $N$, a number which will be large if $h$ is
small, as we envisage.  The approach of this paper avoids 
these difficulties completely by passing to a limiting form of
the problem in which the (negative of the)
log-likelihood expression to be
maximized converges to an integral, the action integral. 
This can be minimized by calculus of variations, so that 
instead of having to do some numerical nonlinear optimization,
we find ourselves having to solve a second-order non-linear
ordinary differential equation  (ODE),
which is a far easier task.  The only issue to be dealt with is
that the ODE is of shooting type, with boundary conditions at
$t=0$ and $t=T$, so that some iterative solution scheme is needed:
the analysis is explained in Section \ref{S2}.

The action integral is only the negative log-likelihood of the 
path in some formal sense, since it involves derivatives of the
paths of the diffusion, and diffusion paths are not differentiable.
Nevertheless, it is a useful heuristic\footnote{
This heuristic leads in a few lines to (non-rigorous) derivations
of the Cameron-Martin-Girsanov change of measure, and to 
large-deviation rate functions for various diffusion asymptotics.}
which leads to a rigorous approximation result for the limiting
form of the maximum-likelihood path as $h \downarrow 0\;$:
see Theorem \ref{thm1}.

This gives us a way to identify effectively the `most likely' path $\jx^*$
of $\jX$ given the path of $\jY$, but we would also like to have some
idea of how the random path of $\jX$ is distributed around
$\jx^*$. This involves a study of the log-likelihood in a 
neighbourhood  of $\jx^*$ which we find is (to leading
order)  quadratic in the perturbation $\jxi = \jx - \jx^*$. Thus
the perturbation $\jxi$ is approximately a Gaussian process, 
whose mean is identically zero, and whose law can be
 precisely characterized by expressing $\jxi$ as the solution of
 a linear stochastic differential equation: see Section \ref{S3}.
 
 Inference on a hidden diffusion intensity for
 a point process is discussed in Section \ref{S4}, and 
 the relationship with SMC methodologies is discussed briefly
 in Section \ref{S5}. Section \ref{conc} concludes.
 
The idea of studying the action of a diffusion path is quite
ancient already, and even as a tool for estimation there is
a literature: see, for example, \cite{bom}, \cite{acost}, \cite{FPRS}.
Markussen \cite{bom} arrives at essentially the same 
identification of the maximum-likelihood path as we do;
the expression derived here for the Gaussian law of the
perturbation is considerably more explicit. The approach of 
Archambeau et al \cite{acost} has superficial similarities, but
is different in major respects; in particular, they use a
minimum relative entropy criterion to identify a `best'
Gaussian approximation to the hidden process, whose
structure appears to require the calculation of 
expectations of non-linear functionals of the path. 
In contrast, the approach followed here requires
only the solution of ordinary differential equations\footnote{
Archambeau et al  work with a more restricted diffusion
dynamic (which they assert can be extended), and a somewhat
different structure for the relation between observations and
the hidden process.}  This is particularly advantageous in 
higher dimensions, as the numerical solution of ordinary 
differential equations still works quite effectively in moderately
large dimension, whereas methods such as particle filtering
struggle in dimension much above about seven, as we shall
discuss in Section \ref{S5}.

\section{The setting.}\label{S1}
Suppose that we have some stochastic
differential equation (SDE) in $\R^d$
\begin{equation}
d\jZ_t = \sigma(t,\jZ_t)\,d\jW_t + \jmu(t,\jZ_t) dt,
\label{dZ}
\end{equation}
where $\jZ$ is partitioned $\jZ = [\jX;\jY]$
and  where $\jX$ is $n$-dimensional,  $\jY$ is
$s$-dimensional, $n+s = d$.  We shall suppose also 
that $\sigma$, $\sigma^{-1}$ and $\jmu$ are in $C^1_b$. 
The problem we address is to estimate the hidden part
$(\jX_t)_{0 \leq t \leq T}$ of $\jZ$ given observations of 
$(\jY_t)_{0 \leq t \leq T}$; the possibility that $\jY$ is non-informative
about $\jX$ is included in the analysis, so what follows applies 
to the distribution of an unobserved diffusion also.

Closely related to \eqref{dZ} is the first-order Euler-Maruyama
difference scheme
\begin{equation}
   d\jz^{(n)}_t = \sigma(\nnt, \jz^{(n)}_{\nnt}) \, d\jW_t +
    \jmu(\nnt, \jz^{(n)}_{\nnt}) \, dt 
    \label{dzn}
\end{equation}
where   $\nnt \equiv 2^{-n} [ 2^n t]$. Despite appearances,
 this can be viewed as a discrete  scheme; the increments of
 $\jz$ have conditionally Gaussian distributions.
  It is known 
 (see, for example, Mao \cite{mao}) that for some constant
 $C$
 \begin{equation}
   E \bigl[\;  \sup_{0 \leq t \leq T} | \jZ_t - \jz^{(n)}_t|^2  \; \bigr]
   \leq C 2^{-n},
   \label{conv1}
\end{equation}
so by the first Borel-Cantelli Lemma we conclude that there is 
almost-sure uniform convergence of the processes 
$\jz^{(n)}$ to the solution $\jZ$ to the SDE.

In the discrete-time approximation \eqref{dzn} to the SDE,
the conditional density of  $\jx(j2^{-n})_{0 \leq j \leq 2^nT}$
given  $\jy(j2^{-n})_{0 \leq j \leq 2^nT}$ can be written down
immediately. The log-likelihood is 
\begin{eqnarray}
\lambda_n(\jx|\jy)  &=&  - \half \sum_{j=0}^{N-1} \frac{1}{h} \;
    \bigl\vert \; \sigma(jh,\jz_{jh})^{-1} \bigl( \; \jz_{jh+h} - \jz_{jh}
    - h \jmu(jh,\jz_{jh}) \;\bigr) \; \bigr\vert^2 - \varphi(\jx_0)
    \nonumber
    \\
    &=&  - \half \sum_{j=0}^{N-1}h \;
    \bigl\vert \; \sigma(jh,\jz_{jh})^{-1} \bigl( \; \frac{\jz_{jh+h} - \jz_{jh}}{h}
    - \jmu(jh,\jz_{jh}) \;\bigr) \; \bigr\vert^2 - \varphi(\jx_0),
    \label{LL1}
\end{eqnarray}
where the prior density of $\jx_0$ is $\exp( - \varphi(\jx))$,
and $h = 2^{-n}$.  Inspecting \eqref{LL1}, we see a difference
quotient of $\jz$ which it is tempting to replace with a derivative
as $n \rightarrow \infty$, leading to the {\em formal} limit
\begin{eqnarray}
\Lambda(\jx|\jy)  &=& -\half  \int_0^T  \bigl\vert \; \sigma(s,\jz_{s})^{-1} \bigl( \; \dot{\jz}_s
    - \jmu(s,\jz_s) \;\bigr) \; \bigr\vert^2 \; ds -\varphi(\jx_0)
    \label{LA1}
    \\
    &\equiv &  -  \int_0^T  \psi(s,\jx_s,\jp_s)\; ds -\varphi(\jx_0)
    \label{LA2}
\end{eqnarray}
where the Riemann sum becomes an integral,
and we write $\jp_s \equiv \dot \jx_s$. This is a 
perfectly sensible functional of a $C^1$ path $\jz$, even though
for a diffusion process the path will not be differentiable.  

Our viewpoint  is that the problem which 
concerns us is to estimate $\jx(j2^{-\nu})_{0 \leq j \leq 2^\nu T}$
 in the {\em discretely-sampled}
model \eqref{dzn}, for some {\em fixed} (quite large) integer
$\nu$,  and that the SDE version of the problem
is to be used to help us in this. In practice, we will only ever
have the observation data $\jy(j2^{-\nu})_{0 \leq j \leq 2^\nu T}$
 in discretely-sampled form (how
would it be stored if not?!) so this is the realistic question.
When we minimize the log-likelihood \eqref{LL1} over $\jx$, 
we only need the values of $\jy$ at the multiples of $h
=2^{-\nu}$; we 
therefore lose no generality in supposing that $\jy$ has
been
 interpolated (by cubic splines, say) to be $C^2$
in all of $[0,T]$.
 This does not of course change \eqref{LL1}, 
but it does mean that the functional $\Lambda$ is meaningfully
defined for any $C^1$ path $(\jx_t)_{0 \leq t \leq T}$. Given this,
we have the following result, where to emphasize again, 
we are considering what happens as $n \geq \nu$ tends to 
infinity, with $\nu$ fixed.

\begin{theorem}\label{thm1}
Suppose that $\sigma$, $\sigma^{-1}$ and $\jmu$ are
$C^1_b$, and that $(\jy_t)_{0 \leq t \leq T}$ is also $C^1$.
Suppose further that 
\begin{equation}
\lim_{|\jx|\rightarrow\infty} \varphi(\jx)
= \infty.
\label{regcon}
\end{equation}
Then 
\begin{equation}
  \lim_{n\rightarrow\infty} \inf_\jx\bigl\lbrace\;
   - \lambda_n(\jx|\jy)\; \bigr\rbrace
  = \inf_\jx\bigl\lbrace\; -\Lambda(\jx|\jy)
  \;\bigr\rbrace.
  \label{t1}
\end{equation}
Assuming that $\Lambda(\cdot|\jy)$ has a unique maximizer
$\jx^*$,
and that $\jx_n$ is a maximizer of $\lambda_n(\cdot|\jy)$, then 
$\jx_n \rightarrow \jx^*$ uniformly.
\end{theorem}

\medskip
\noindent{\sc Proof.}  See Appendix \ref{app}.

\medskip\noindent
The usefulness of Theorem \ref{thm1} is that it shows
that a minimizer of $-\lambda_n(\cdot|\jy)$ is very close to 
the (assumed unique) minimizer of $-\Lambda(\cdot|\jy)$;
but this minimizer can be found by calculus of variations
without resort to some high-dimensional non-linear
optimizer.

Notice that although we make a discretization error when we 
replace the SDE  \eqref{dZ} with the Euler-Maruyama scheme
\eqref{dzn}, in most examples of practical interest the magnitude
of this discretization error will be insignificant compared to the 
observation error that we are attempting to see past.

\section{Finding the least-action path.}\label{S2}
Theorem \ref{thm1} shows that asymptotically as the step
size $h=2^{-n}$  tends to zero, the maximum-likelihood
estimate of the unobserved path converges to the maximizer
of the continuous-time analogue $\Lambda$.  Resorting now
to calculus of variation techniques leads us to the following
central result.

\begin{theorem}\label{thm2}
The path $(\jx,\jp)$ which minimizes the action functional
\begin{equation}
-\Lambda(\jx|\jy) =    \int_0^T  \psi(s,\jx_s,\jp_s)\; ds +\varphi(\jx_0)
\label{AF}
\end{equation}
satisfies the following ODE with boundary conditions:
\begin{center}
\setlength{\fboxrule}{0.5pt}
\setlength{\fboxsep}{1mm}
\fbox{
\parbox[c]{0.8\linewidth}{
\begin{eqnarray}
0 &=& D_{\jp_j}\psi(0,\jx_0,\jp_0) - D_{\jx_j}\varphi(\jx_0)
\label{bc0}
\\
0&=& D_{\jx_j}\psi - D_tD_{\jp_j}\psi - \dot \jx_k D_{\jp_j}D_{\jx_k}\psi
-\dot \jp_k D_{\jp_j}D_{\jp_k} \psi
\label{ode}
\\
0 &=& D_{\jp_j}\psi(T,\jx_T,\jp_T)
\label{bcT}  
\end{eqnarray}
}  }
 \end{center} 
\end{theorem}

\medskip\noindent
{\sc Proof.}
The negative 
 log-likelihood to be minimized is
\begin{equation}
 \varphi(\jx_0) +  \int_0^T \psi(t,\jx_t,\dot \jx_t) \; dt
\equiv  \varphi(\jx_0) +  \int_0^T \psi(t,\jx_t,\jp_t) \; dt,
\label{LL}
\end{equation}
where we write $\jp \equiv \dot \jx$.  This we attack by calculus of variations; 
if we have found the optimal $\jx$, then any perturbation to $\jx + \jxi$ 
must to leading order make zero change to the objective. 
Writing down the first-order change and  integrating by
parts gives 
\begin{eqnarray}
0 &=& \jxi_0 D\varphi(\jx_0) + \int_0^T \bigl\lbrace \jxi_t \cdot D_\jx\psi
	+ \dot\jxi \cdot D_\jp\psi \bigr\rbrace \; dt
	\label{FOC}
\\
&=& \jxi_0 D\varphi(\jx_0) +\bigl[ \jxi^j_t D_{\jp_j}\psi \bigr]_0^T
+ \int_0^T  \jxi^j_t  \bigl\lbrace
D_{\jx_j}\psi - D_tD_{\jp_j}\psi - \dot \jx_k D_{\jp_j}D_{\jx_k}\psi
-\dot \jp_k D_{\jp_j}D_{\jp_k} \psi
\bigr\rbrace \; dt.
\nonumber
\end{eqnarray}
Since $\jxi$ is arbitrary, we deduce the conditions
\eqref{bc0}, \eqref{ode}, \eqref{bcT}  for optimality.
\hfill$\square$ 
 
 \bigbreak
 \noindent{\sc Remarks.}
 (i) 
 We have found a (generally non-linear)  first-order ODE 
\eqref{ode} for $(\jx,\jp)$, with initial condition \eqref{bc0} and 
terminal condition \eqref{bcT}. This will generally have a unique 
solution, though the `shooting' nature of the ODE is rather clumsy
in practice.  

\noindent
(ii)  We expect that this methodology will be advantageous in 
higher-dimensional problems;  algorithms for  solving ODEs in high
dimension tend to degrade less rapidly than (for example)
gradient optimization methods, or particle filtering.  

\noindent
(iii)  If the time horizon $T$ is too large, it may be that the
shooting ODE is not able to identify the initial and terminal 
conditions with sufficient accuracy.

\medskip\noindent
To illustrate the methodology in action, we present here some
simple examples.

\begin{figure}[H]
\begin{center}
\includegraphics[height=16cm,width=12cm,angle=-90]
{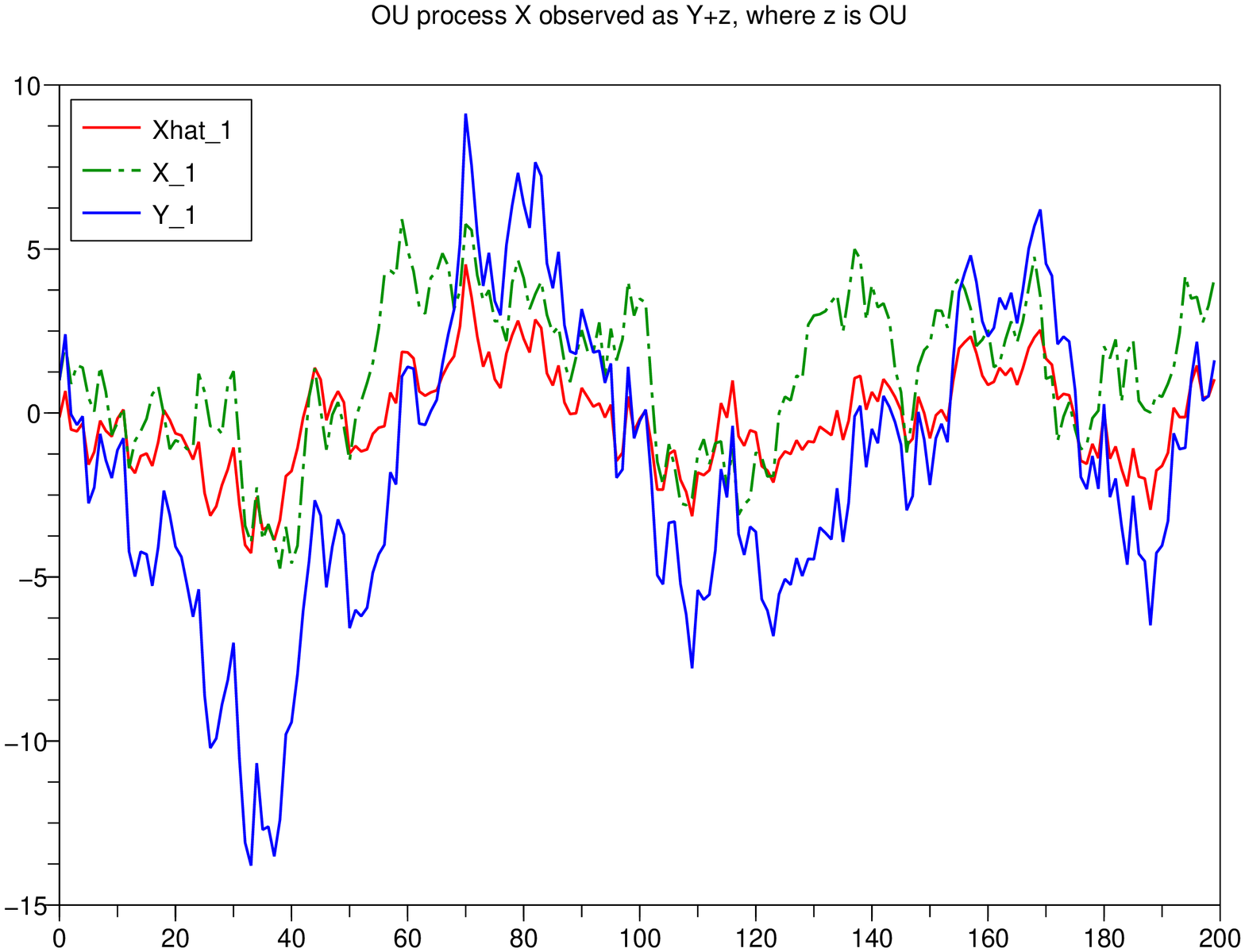} 
\end{center}
\caption{Result of least-action filtering.
}
\label{EG1pic}
\end{figure}

\medskip\noindent
{\bf Example 1.}  Here we take independent  one-dimensional
Ornstein-Uhlenbeck  processes $X$, $y$
\begin{eqnarray*}
dX &=& \sigma_X dW - \beta_X X dt
\\
dy &=& \sigma_y dW' - \beta_y Y dt
\end{eqnarray*}
where $\sigma_X = 1.053$, $\sigma_y =1.0127 $,
 $\beta_X = 0.1054$
and $\beta_y = 0.0253$.   We 
observe $Y = X+y$.    Then $\jZ = [X;Y]$ solves
  \begin{eqnarray*}
d\jZ &=& 
\biggl( \begin{array}{cc}
\sigma_X & 0 \\ \sigma_X & \sigma_y
\end{array}
\biggr)
	\biggl( \begin{array}{c}
dW \\dW'
\end{array}
\biggr)  + \biggl( \begin{array}{cc}
-\beta_X & 0 \\ -\beta_X +\beta_y& -\beta_y
\end{array}
\biggr) \biggl( \begin{array}{c}
X\\Y
\end{array}
\biggr)  dt
\\
&\equiv & 
\sigma
	\biggl( \begin{array}{c}
dW \\dW'
\end{array}
\biggr)  + A\biggl( \begin{array}{c}
X\\Y
\end{array}
\biggr)  dt,
\nonumber
\end{eqnarray*}
which is a simple linear SDE.  For this example, 
\begin{equation}
\psi(t,x,p) = \frac{1}{2}
\biggl( \, \biggl( \begin{array}{c}
p \\ \dot Y_t  \end{array}  \biggr)  - A \biggl( \begin{array}{c}
x \\ Y_t  \end{array}  \biggr) \; \biggr)^T
	q
\biggl( \, \biggl( \begin{array}{c}
p \\ \dot Y_t  \end{array}  \biggr)  - A \biggl( \begin{array}{c}
x \\ Y_t  \end{array}  \biggr) \; \biggr).
\nonumber
\end{equation}
A sample path of the SDE was simulated, and the results are
displayed in Figure \ref{EG1pic}. The true path is dashed in green,
the noisy observation is in blue, and the least-action path is
in red.

\begin{figure}[H]
\begin{center}
\includegraphics[height=16cm,width=12cm,angle=-90]
{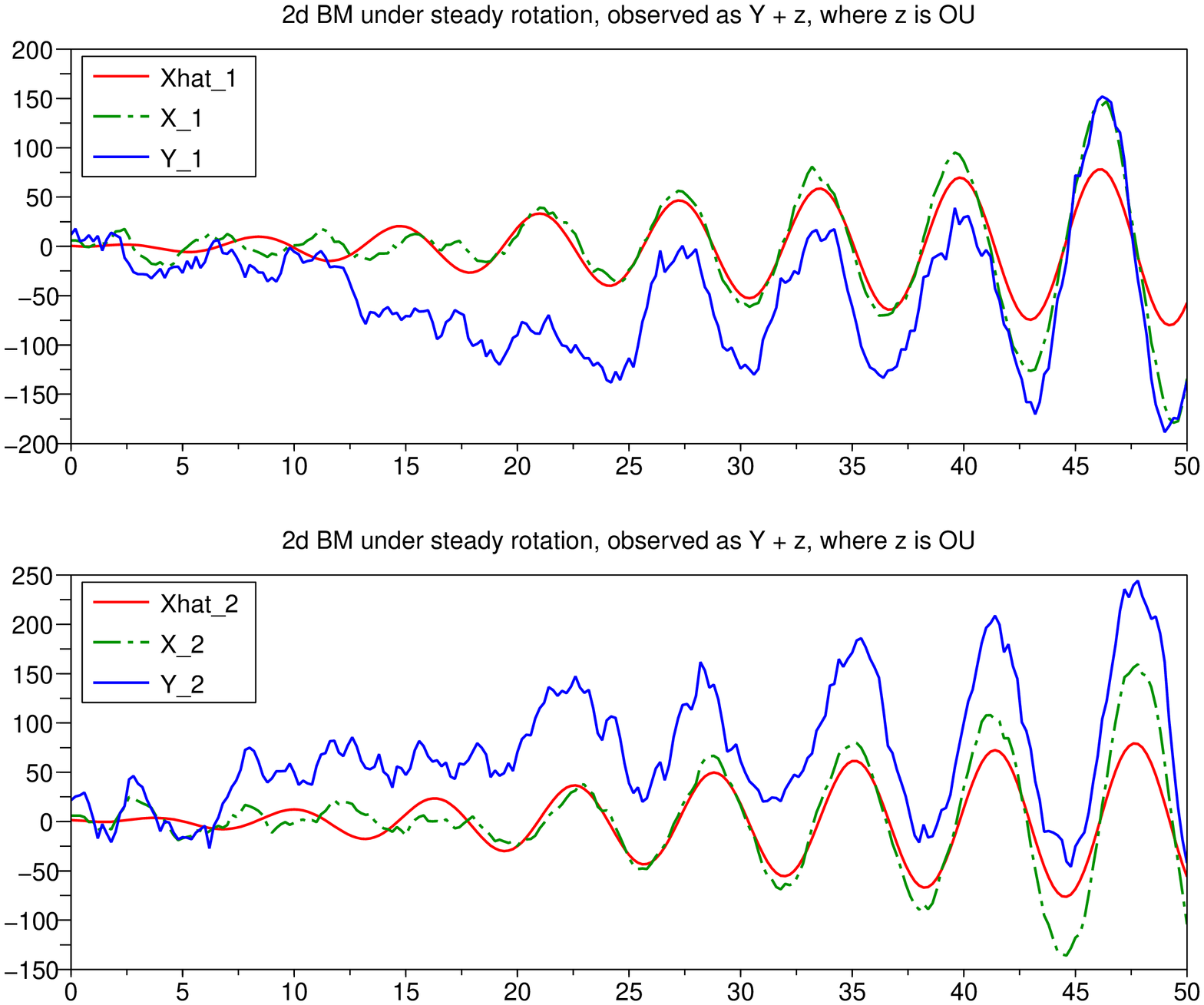} 
\end{center}
\caption{Result of least-action filtering.
}
\label{EG2pic}
\end{figure}

\medskip\noindent
{\bf Example 2.}
This time we have a two-dimensional linear example, with
$\jX$ satisfying
\[
    d\jX_t = d\jW_t + \begin{pmatrix}
0 & -1 \\ 
1 & 0
\end{pmatrix} \jX_t \; dt \equiv d\jW_t + A_0 \jX_t \; dt,
\]
and $\jY = \jX+\jy$, where $\jy$ is an independent OU process
\[
    d\jy_t = \sigma d\jw_t - \lambda \jy_t \; dt,
\]
with $\lambda = 0.005$, $\sigma = 20$.
The results of the least-action analysis are displayed in 
Figure \ref{EG2pic}. Once again, the estimates are quite
impressive, but since the underlying dynamics are a perturbation
of uniform motion in a circle, this is perhaps not so surprising.

\begin{figure}[H]
\begin{center}
\includegraphics[height=18cm,width=13cm,angle=-90]
{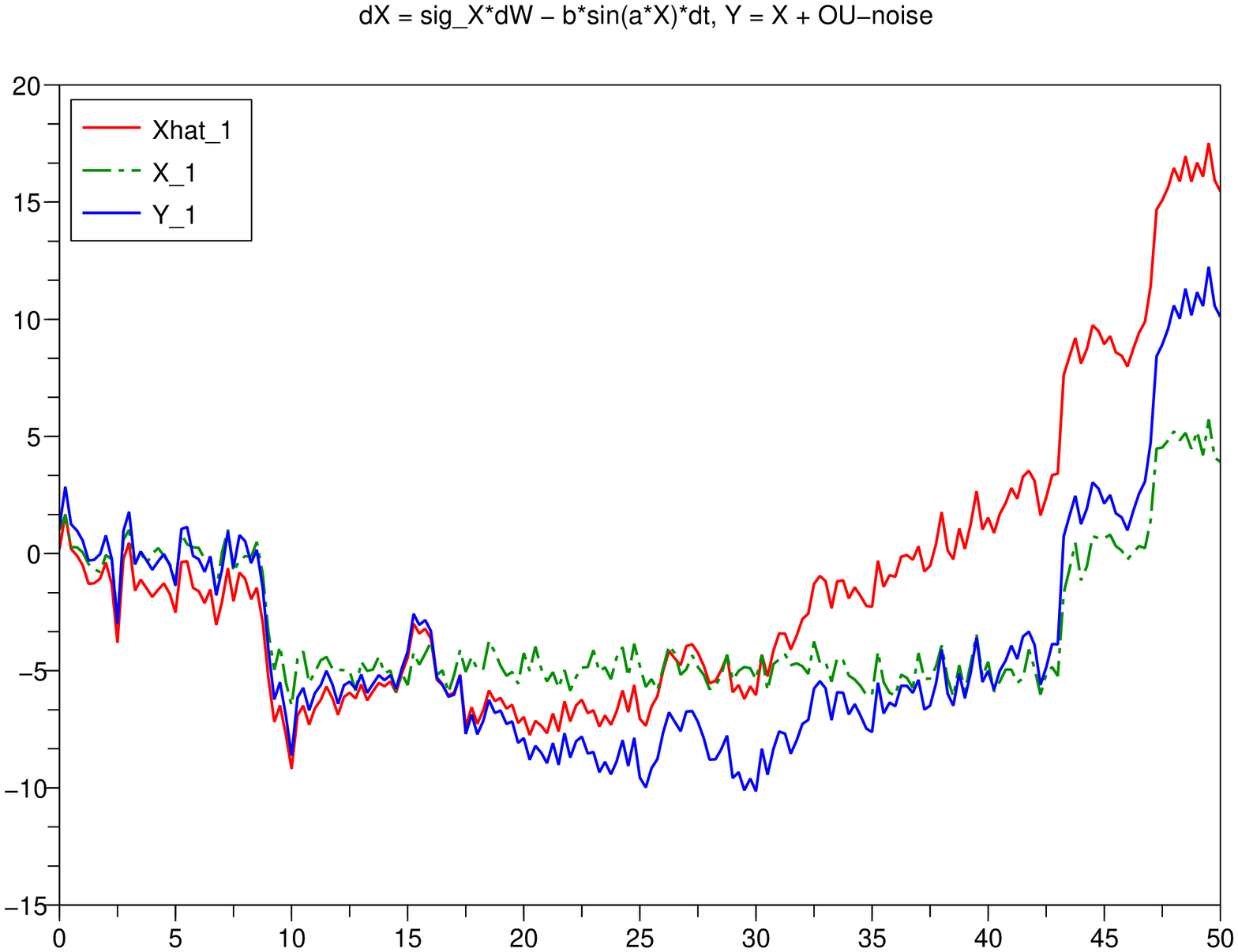} 
\end{center}
\caption{Result of least-action filtering.
}
\label{EG3pic}
\end{figure}

\medskip\noindent
{\bf Example 3.}  This is another one-dimensional example, but 
this time non-linear. We have dynamics
\[
    dX_t =\sigma_X  dW_t -b\sin(aX_t) \; dt
\]
for $X$, where $\sigma_X = 3$,  $b=12$ and $a = 2\pi/5$.
Thus the drift tries to keep the diffusion near to multiples of 5.
The observation $Y$ is as before of the form
 $Y = X+y$ where $y$ is an independent OU process
\[
    dy_t = dW'_t - \lambda y_t \; dt
\]
with $\lambda = 0.05$.  The results of the analysis are shown 
in Figure \ref{EG3pic}.  As can be seen, the hidden Markov process
stays close to multiples of 5, occasionally moving from one value to 
a neighbouring value, and as the observational error gets larger, the 
filtering  performs less well.

\begin{figure}[H]
\begin{center}
\includegraphics[height=17cm,width=12cm,angle=-90]
{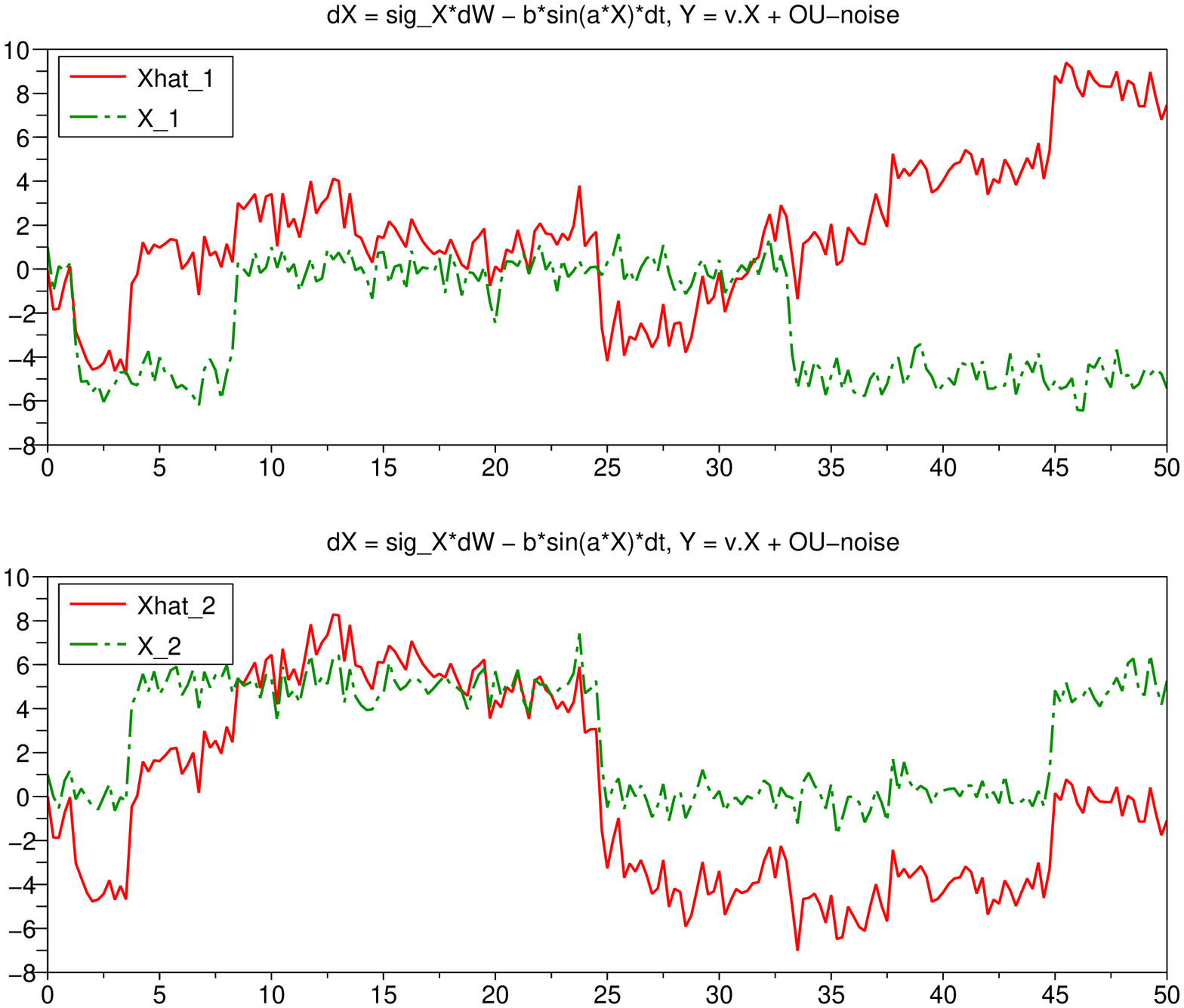} 
\end{center}
\caption{Result of least-action filtering.
}
\label{EGpic4}
\end{figure}

\medskip\noindent
{\bf Example 4.}   The final example is a more demanding test of
the methodology. Here, we take a two-dimensional non-linear
dynamic for $\jX$
\[
    d\jX_t = \Sigma d\jW_t -b\sin(a\jX_t) \; dt
\]
where 
\begin{equation}
\Sigma^2 = \begin{pmatrix}
0.9 & 0.27 \\ 
0.27 & 0.9
\end{pmatrix} 
\end{equation}
$a, b$ as for the previous example.  The observation is 
{\em univariate}, and takes the form
 $Y = \jv \cdot \jX+y$ where $y$ is an independent OU process
\[
    dy_t = \sigma dw_t - \lambda y_t \; dt
\]
where $\jv = (1,2)$,  $\lambda = 0.05$, and $\sigma = 1$.
  It would be indeed remarkable
if the methodology was able to unscramble the two hidden components
of the diffusion from observation of just one component, and the
results are shown in Fig \ref{EGpic4}; sometimes the estimate is 
close to the true values, sometime less so. But it should be
understood that this is not a limitation of the methodology, which
is after all discovering the maximum-likelihood estimator of the
hidden process; the relatively poor recovery of the hidden path
is because we have a small signal-to-noise ratio for this example.

\section{Second-order analysis.}\label{S3}
The first  problem we focused on is maximizing the log-likelihood
as given for the discretized problem by  \eqref{LL1}. 
The unknown in this situation is the sequence
 $\sX \equiv (\jx_{j2^{-\nu}})_{0 \leq
j\leq 2^\nu T}$, and as in classical asymptotic ML  theory, 
having identified the ML estimator $\hat\sX$ of the unknown $\sX$, 
we can enquire about the distribution of $\sX$ about 
$\hat\sX$. This amounts to understanding the second derivative
of the log-likelihood at the MLE.  Given suitable
regularity of the likelihood, a Taylor expansion shows that
the distribution of $\sX$ around $\hat \sX$ will be 
approximately Gaussian, with covariance equal to 
the second derivative of the log-likelihood.
 However, we have seen that
there are considerable methodological advantages in 
passing from the discrete problem \eqref{LL1} to the
continuous analogue \eqref{LA2}, and these advantages apply
also at the level of the second-order terms in the expansion
of the log-likelihood, as we shall now see.

\begin{theorem}\label{thm3}
Suppose that $\jx^*$ is a local minimizer of the functional
$\Lambda(\cdot|\jy)$, and define the matrix-valued $n \times n$
functions of time 
\begin{equation}
A^{ij}_t = D_{\jx_i}D_{\jx_j} \psi,
\qquad
B^{ij}_t = D_{\jx_i}D_{\jp_j} \psi,
\qquad
q^{ij}_t = D_{\jp_i}D_{\jp_j} \psi,
\label{ABqdef}
\end{equation}
evaluated along the path $\jx^*$. Then:
\begin{itemize}
\item[(i)] the ODE
\begin{equation}
A_t + \dot\theta_t = K_t^T q_t K_t = (B_t + \theta_t) q_t^{-1}
(B^T_t + \theta_t),
\qquad \theta_T = 0
\label{thetaODE}
\end{equation}
has a unique solution;
\item[(ii)] writing  $\jx = \jx^* + \jxi$, we have that
\begin{equation}
-\Lambda(\jx|\jy) + \Lambda(\jx^*|\jy)
= Q_0(\jxi) + o( \int_0^T (|\jxi_s|^2 + |\dot\jxi_s|^2)\; ds),
\label{La2}
\end{equation}
where
\begin{equation}
 Q_0(\jxi) =  \half \jxi_0 \cdot \bigl\lbrace\,
 D^2\varphi(\jx^*_0) + \theta_0\,\bigr\rbrace \jxi_0 
 + \int_0^T  \half (\dot\jxi_t + K_t\jxi_t) \cdot q_t (\dot\jxi_t + K_t\jxi_t)
 \; dt,
 \label{Q0def}
\end{equation}
and where
\begin{equation}
K_t = q_t^{-1}(B^T_t + \theta_t).
\label{Kdef}
\end{equation}
\end{itemize}
\end{theorem}

\medskip\noindent
{\sc Remarks.}  (i)  If we write $\dot w_t =  q_t^{1/2} (\dot\jxi_t
+ K_t \jxi_t)$, then the integral term in \eqref{Q0def} 
takes the form $\int_0^T |\dot w_t|^2\; dt$, which is the 
action integral for standard Brownian motion. Thus informally 
what we learn from Theorem \ref{thm3} is that the perturbation
$\jxi$ `looks like' a continuous zero-mean Gaussian process obtained
by solving the linear SDE
\begin{equation}
d\jxi_t = - K_t \jxi_t \, dt + q_t^{-1/2} \, dW_t,
\label{dxi}
\end{equation}
and started with initial precision $D^2\varphi(\jx^*_0) + \theta_0.$
 The covariance of $\jxi$ can thus  be  obtained from an
It\^o expansion of $\jxi\jxi^T$:
\begin{equation}
d \jxi\jxi^T \doteq ( - K_t \jxi_t\jxi_t^T - \jxi_t \jxi_t^TK_t^T + q_t^{-1}) dt,
 \label{dxixi}
\end{equation}
where $\doteq$ signifies that the two sides differ by a (local) martingale. 
Hence
\begin{equation}
\dot V_t = - K_t V_t - V_t K_t^T + q_t^{-1}
\label{dV}
\end{equation}
and this allows us to calculate the covariance at time $t$,
again by solving an ODE.

\medskip\noindent
(ii) The ODE \eqref{thetaODE} for $\theta$ is quadratic, so there
is no general result which guarantees that it has a solution; the 
solution may explode. However, in this situation we can be sure
that it will not, because $\Lambda(\cdot|\jy)$ is minimized at $\jx^*$, 
as we shall see in the proof.

\medskip\noindent
(iii) It has to be admitted that the relationship between 
Theorem \ref{thm3} and the original discretized problem 
is somewhat tenuous. With effort it would no doubt be possible
to establish an analogue of Theorem \ref{thm1} for the 
second-order effect, but it is harder to see what the use of such
a result would be. The way we envisage using Theorem \ref{thm3}
would be to generate an  approximate distribution for the posterior
of $\sX$ given $\jy$, which could be checked numerically against
a full posterior\footnote{It would be rare that this could be 
found in closed form, so we would typically need to resort to 
particle filtering to make a numerical approximation.}, and which
could be used to give approximate moments of the posterior 
law of $\sX$.

\medskip\noindent
{\sc Proof of Theorem \ref{thm3}.}
Let us begin by assuming the first statement (i) of the theorem
and showing how this leads to the more interesting statement (ii).
We will return to prove (i) subsequently.

If we go back to the action functional  \eqref{AF} and consider 
expanding around the local minimizer
$\jx^*$ by perturbing to $\jx^* + \jxi$ for small $\jxi$, the second-order
contribution we get is 
\begin{eqnarray}
Q(\jxi) & \equiv & \half D_{\jx_i}D_{\jx_j} \varphi(\jx^*_0) \jxi^i_0 \jxi^j_0
+ \int_0^T \bigl\lbrace
\half \jxi^i\jxi^j D_{\jx_i}D_{\jx_j} \psi  + \jxi^i \dot\jxi^j D_{\jp_j}
D_{\jx_i} \psi + \half \dot\jxi^i \dot\jxi^j D_{\jp_i}D_{\jp_j} \psi
\big\rbrace \; dt
\nonumber
\\
&\equiv & \half D_{\jx_i}D_{\jx_j} \varphi(\jx^*_0) \jxi^i_0 \jxi^j_0
+ \int_0^T \bigl\lbrace
\half \jxi^i_t  A^{ij}_t\jxi^j_t  + \jxi^i_t  B^{ij}_t \dot\jxi^j_t 
 + \half \dot\jxi^i_t   \,q^{ij}_t \dot\jxi^j_t
\big\rbrace \; dt
\label{Qdef}
\end{eqnarray}
where 
the matrix-valued functions of time
$A$, $B$ and $q$ are defined by \eqref{ABqdef}.
This quadratic functional of $\jxi$ characterizes the (approximate) Gaussian
distribution of the perturbation.  Let us notice that because 
$\jx^*$ is assumed to be a minimizer of the action functional, this
quadratic functional of $\jxi$ must be non-negative. 

Now suppose that we have some $C^1$ symmetric-matrix-valued
function of time, $\theta$, such that $\theta_T=0$.  Then we may write
\begin{eqnarray}
Q(\jxi) &= & Q(\jxi) + \half \jxi_0 \cdot \theta_0 \jxi_0
                          + \bigl[  \half \jxi_t \cdot \theta_t \jxi_t \bigr]_0^T
                          \nonumber
\\
  &=& \half \jxi_0 \cdot (D^2\varphi(\jx^*_0) + \theta_0) \jxi_0
  \nonumber
  \\
  &&\qquad\qquad
  +  \int_0^T \bigl\lbrace
\half \jxi_t  \cdot A_t\jxi_t  + \jxi_t \cdot  B_t \dot\jxi_t 
 + \half \dot\jxi_t  \cdot \,q_t \dot\jxi_t
 + \half \jxi_t \cdot \dot\theta_t \jxi_t + \jxi_t \cdot \theta_t \dot\jxi_t
\big\rbrace \; dt.
\label{Q1}
\end{eqnarray}
The quadratic form inside the integral can be expressed as  
\begin{equation}
\half \dot\jxi_t \cdot q_t \dot\jxi_t + \jxi_t\cdot(B_t + \theta_t)\dot\jxi_t
+ \half \jxi_t \cdot( A_t + \dot\theta_t) \jxi_t
= \half (\dot\jxi_t + K_t\jxi_t) \cdot q_t (\dot\jxi_t + K_t\jxi_t),
\label{CTS}
\end{equation}
where $K_t \equiv q_t^{-1}(B^T_t + \theta_t)$,
{\em provided}
\begin{equation}
A_t + \dot\theta_t = K_t^T q_t K_t = (B_t + \theta_t) q_t^{-1}
(B^T_t + \theta_t),
\label{eq6}
\end{equation}
that is, provided  $\theta$ solves the ODE \eqref{thetaODE}.  Therefore
we have shown that the second statement of the theorem follows once we have
established the first.

\medskip

We now turn to  the issue of the 
existence\footnote{Uniqueness of the solution is
not a problem, because the right-hand side of the ODE \eqref{thetaODE}
is locally Lipschitz in $\theta$, and once we know that there exists
a solution this is enough to guarantee uniqueness.
} of a solution to \eqref{thetaODE}. 
 The proof that the quadratic
ODE \eqref{thetaODE} has a solution exploits the only piece
of information we have about the quadratic form $Q$, namely 
that it is {\em non-negative}, since $\jx^*$ maximized $\Lambda(\jx|\jy)$.
We may therefore consider the optimal control problem 
\begin{equation}
G(t,\jx) \equiv \inf \biggl\lbrace \;
\int_t^T  \bigl\lbrace
\half \jxi_s  \cdot A_s\jxi_s  + \jxi_s \cdot  B_s \dot\jxi_s 
 + \half \dot\jxi_s  \cdot \,q_s\,  \dot\jxi_s
\big\rbrace \; ds : \jxi_t = \jx
\;\biggr\rbrace
\label{Gdef}
\end{equation}
which is well posed, and the solution $G$ has a quadratic form:
\begin{equation}
 G(t,\jx) = \half \jx\cdot H_t \jx,
 \label{Fform}
\end{equation}
which is centred, since clearly $G(t,0)= 0$ for all $t$.
Since $G$ is the value function, we shall have that
\begin{equation}
 h(t) \equiv   \int_0^t  \bigl\lbrace
\half \jxi_s  \cdot A_s\jxi_s  + \jxi_s \cdot  B_s \dot\jxi_s 
 + \half \dot\jxi_s  \cdot \,q_s\,  \dot\jxi_s
\big\rbrace \; ds  +G(t,\jxi_t)
\label{hdef}
\end{equation}
is non-decreasing whatever $\jxi$ we choose, and will be
constant if we choose the optimal $\jxi$.  Differentiating 
\eqref{hdef} gives us 
\begin{equation}
\dot h_t = \half \jxi_t  \cdot A_t\jxi_t  + \jxi_t \cdot  B_t \dot\jxi_t 
 + \half \dot\jxi_t  \cdot \,q_t\,  \dot\jxi_t+
 \half \jxi_t \cdot\dot H_t \jxi_t
 + \jxi_t  H_t \dot \jxi_t.
 \label{hdot}
\end{equation}
Minimizing over $\dot\jxi_t$ and equating to zero to find
the optimal $\dot\jxi_t$ gives us
\begin{equation}
 \dot\jxi_t  = - q_t^{-1} \bigl(\, B_t^T \jxi_t + H_t \jxi_t
  \,\bigr),
  \label{xidot}
\end{equation}
so that the minimized value of $\dot h_t$  becomes
\begin{equation*}
 \half \jxi_t  \cdot A_t\jxi_t  +  \half \jxi_t  \cdot\dot H_t \jxi_t
 - \half \bigl(\, B_t^T \jxi_t + H_t\jxi_t
  \,\bigr)\cdot q_t^{-1} \bigl(\, B_t^T \jxi_t + H_t\jxi_t
  \,\bigr).
\end{equation*}
At optimality, this must be zero; so
\begin{equation}
0 = A_t + \dot H_t - (B_t + H_t)\, q_t^{-1}(B_t^T + H_t).
\label{q1}
\end{equation}
Evidently from the definition we have $G(T,\cdot) \equiv 0$,
so that $H$ solves \eqref{thetaODE} with the zero boundary
condition at $t=T$. In other words, we have found $\theta$,
and $\theta=H$.  Notice in particular that the equation
\eqref{xidot} becomes $\dot\jxi_t = - K_t \jxi_t$.

\hfill$\square$

\subsection*{Numerical aspects.}
Theorem \ref{thm3} characterizes the behaviour of the 
path $\jx$ around the local minimizer $\jx^*$, but in any given 
example, this is not the end of the story. The point is that
when we seek $\jx^*$ we resort to numerical means\footnote{All
the calculations for this paper were performed in Scilab
{\tt www.scilab.org} which is a free and very versatile
package able to perform a wide range of numerical and
graphical tasks.} to 
solve the ODE \eqref{ode} subject to the boundary conditions 
\eqref{bc0}, \eqref{bcT}, and it may be that the solution 
found is not the global minimizer. It could be that the solution
obtained is only a local minimizer; it could be that the solution 
found is a only stationary point of $\Lambda(\cdot|\jy)$; or it could be
that the numerical method has got into difficulties and what it
returns is not a solution. An obvious approach to this problem 
is simply to try numerical solution of \eqref{thetaODE} back in 
time from initial value $\theta_T=0$; however, if the ODE solver
fails, usually the entire calculation stops, and no attempt to 
continue is possible.  We can avoid this issue if we transform
the first-order non-linear Riccati ODE  into a second-order 
linear ODE. The following result gives all we need.

\begin{proposition}\label{prop1}
Suppose that $\jx^*$ is a solution to \eqref{bc0}, \eqref{ode},
\eqref{bcT}, and that $A$, $B$ and $q$ are defined in terms of 
$\jx^*$ by \eqref{ABqdef}. Then:
\begin{itemize}
\item[(i)]  If the ODE \eqref{thetaODE} has a unique
solution, then  the  $n\times n$-matrix-valued
function $F_t$ defined as the unique solution to the first-order
linear ODE
\begin{equation}
   \dot{F}_t = - K_t F_t\equiv -q_t^{-1}
   (B_t^T+\theta_t) F_t, \qquad F_T = I
   \label{Fdef}
\end{equation}
 solves the second-order linear ODE
\begin{equation}
0 = (A_t - \dot B_t^T)F_t + (B_t - B^T_t - \dot{q}_t) \dot F_t - q_t \ddot F_t
\label{FODE}
\end{equation} 
with boundary conditions
\begin{equation}
 F_T = I, \qquad B_T^T F_T + q_T \dot{F}_T = 0.
 \label{FBC}
\end{equation}
\item[(ii)]   If the (unique) solution to \eqref{FODE} with 
boundary conditions \eqref{FBC} is non-singular for all time, 
then $\theta$ defined by
\begin{equation}
     B^T_t + \theta_t = - q_t \dot{F}_t F_t^{-1}
     \label{thF}
\end{equation}
 solves \eqref{thetaODE}, and is the unique solution.
\item[(iii)]   If the (unique) solution to \eqref{FODE} with 
boundary conditions \eqref{FBC} fails to be non-singular
for all time, then 
$\jx^*$ cannot be a local minimizer of $\Lambda(\cdot|\jy)$.
\end{itemize}
\end{proposition}

\bigbreak\noindent{\sc Proof.}
(i) Suppose that $\theta$ is the unique solution to \eqref{thetaODE},
 and use this to define a function $F$ using \eqref{Fdef} with
  conditions  \eqref{FBC} at $T$. Since \eqref{Fdef} is a linear
  ODE with bounded continuous coefficients, it has a unique solution.
  Multiplying \eqref{Fdef} on the left by $q_t$ and differentiating
  leads to 
\begin{eqnarray}
-q \ddot{F} - \dot{q} \dot{F} &=&
    (\dot{B}^T + \dot{\theta}) F + (B^T+\theta)\dot F
    \nonumber
    \\
    &=&   (\dot{B}^T-A +(B+\theta)  \, q^{-1}(B^T+\theta) ) F
     + (B^T+\theta) \dot F
     \label{use18}
     \\
     &=&  (\dot{B}^T-A)F - (B+\theta)\dot{F}+ (B^T+\theta) \dot F
     \nonumber
     \\
     &=&   (\dot{B}^T-A)F - (B-B^T)\dot{F}
     \nonumber
\end{eqnarray}
which is the ODE \eqref{FODE}, as required. Here, we have used
\eqref{thetaODE} at line \eqref{use18}.

\medskip\noindent
(ii)  Using \eqref{thF}
to define a function $\theta$ of time, we see from \eqref{FBC} that
$\theta_T=0$, so it remains to check that $\theta$ so defined
satisfies \eqref{thetaODE}.  Multiplying on the right by $F$ and
differentiating \eqref{thF} gives us
\begin{equation}
(\dot{B^T} + \dot{\theta} )F + (B^T+\theta)\dot{F}
=   - \dot{q} \dot{F}  - q \ddot{F}
 = (\dot{B^T} - A) F+(B^T-B)\dot{F}
 \label{Feq}
\end{equation}
leading to the equation  
\begin{equation}
(A+\dot{\theta}) F   = -(B+\theta)\dot{F}
=(B+\theta)  \, q^{-1}(B^T+\theta)F.
\end{equation}
Since $F$ is assumed to be invertible for all $t$, we can 
clear out the common factor  of $F$ on each side and recover
the ODE \eqref{thetaODE} for $\theta$.

\medskip\noindent
(iii)     
Suppose that $\jx^*$ were a local minimizer of $\Lambda(\cdot|\jy)$;
then by Theorem \ref{thm2}, the ODE \eqref{thetaODE} has a unique
solution.  We may therefore define a $n\times n$ matrix-valued
 function $f_t$
of time to solve
\begin{equation*}
\dot f_t = -K_t f_t, \qquad f_T=I
\end{equation*}
where $K_t = q_t^{-1}(B_t^T + \theta_t)$.  Notice that $K$ is bounded
on $[0,T]$, so we may also define the  $n\times n$ matrix-valued function
$\varphi_t$ by
\[
\dot \varphi_t = \varphi_t K_t,\qquad \varphi_T=I.
\]
Calculus shows that $\varphi_t f_t$ is constant, and therefore
equal to $I$; in particular, $f_t$ is always invertible.
According to the first statement of the Proposition, $f$ solves
the ODE \eqref{FODE} subject to the boundary conditions \eqref{FBC},
which has a unique solution. But by hypothesis, this solution 
becomes singular at some time in $[0,T]$, which is a contradiction. 
The conclusion follows.

\hfill$\square$

\bigbreak
\noindent
{\sc Remarks.}  The way Proposition \ref{prop1} gets used in 
practice is that we firstly identify what we think may be the 
action-minimizing path $\jx^*$ and then we  solve the 
second-order linear ODE \eqref{FODE} with the boundary
conditions \eqref{FBC}. There is never any problem with this
ODE, since it is linear with bounded continuous coefficients.
Having solved for $F$, we now calculate the determinant of 
$F_t$ along the path of $F$.  If this ever vanishes, it indicates
that  $\jx^*$ was not a local minimizer  of $Q_0$.


\section{An extension.}\label{S4}
We have so far considered a situation where a diffusion
$\jX$ is observed through some continuous process $\jY$ of
observations, and we are required to estimate $\jX$. Another
quite natural situation we might want to consider would 
be where the hidden process $\jX$ is some positive diffusion, 
which acts as the intensity process of some observed point 
process.  So suppose that there is some non-negative
diffusion $x$ satisfying
\begin{equation}
dx_t =    \sigma(t,x_t) \, dW_t + \mu(t,x_t )\, dt
\label{dxdef}
\end{equation}
which serves as the stochastic intensity of a counting process $N$. The
times $0 < \tau_1 < \tau_2<\ldots < \tau_{N_T} < T$ of events are observed,
and we have to filter $x$ from these observations.  In this case, 
the action integral ( = -log-likelihood) to be minimised is just
\begin{equation}
 \varphi(x_0) + \int_0^T \psi(s,x_s,\dot x_s) \; ds + 
 \int_0^T x_s \; ds
- \sum_{i=1}^n \log x_{\tau_i},
\label{LL2}
\end{equation}
where we have abbreviated $n \equiv N_T$, and $\psi$ is as before
\begin{equation}
\psi(s,x,p) = (p-\mu(s,x))^2/2\sigma(s,x)^2.
\end{equation}
Once again, we consider 
a small perturbation $\xi$ away from the optimal $x$, and equate the
first-order change to zero. Assuming for simplicity
that $\sigma$ and $\mu$ are independent of time gives us\footnote{We 
write $\tau_0 = 0$, $\tau_{n+1} = T$.} 
\begin{eqnarray*}
0 &=& \varphi'(x_0)\xi_0 + \int_0^T \,  \{
                        \psi_x \xi + \psi_p \dot\xi + \xi \} \; dt 
                                - \sum_{i=1}^n \frac{\xi(\tau_i)}{x(\tau_i)}
\\
&=& \varphi'(x_0)\xi_0 + \sum_{j=0}^n\int_{\tau_j}^{\tau_{j+1}} \,  \{
                        \psi_x \xi + \psi_p \dot\xi + \xi \} \; dt 
                                - \sum_{i=1}^n \frac{\xi(\tau_i)}{x(\tau_i)}
\\
&=& \varphi'(x_0)\xi_0 + \sum_{j=0}^n \biggl[ \;
\int_{\tau_j}^{\tau_{j+1}} \,  \{
                        \psi_x -\psi_{px} \dot x- \psi_{pp}\dot p +1  \} \xi \; dt 
				+ \Bigl[ \xi_t \psi_p(x_t,\dot x_t) \Bigr]_{\tau_j}^{\tau_{j+1}}
\;\biggr]
                                - \sum_{i=1}^n \frac{\xi(\tau_i)}{x(\tau_i)}.
\end{eqnarray*}
In order for this to be zero whatever perturbation $\xi$ is used, we have to have
a number of conditions:
\begin{eqnarray}
0&=&  \psi_x -\psi_{px} \dot x- \psi_{pp}\dot p +1\quad \textrm{ in each interval
                             $(\tau_j,\tau_{j+1})$ \; ;}
\label{c1}
\\
0 &=& \varphi'(x_0) - \psi_p(x_0,p_0)\;  ;
\label{c2}
\\
0 &=& -\psi_p(x_{\tau_i},p_{\tau_i+}) + \psi_p(x_{\tau_i},p_{\tau_i-}) 
-x(\tau_i)^{-1} \;  ;
\label{c3}
\\
0 &=& \psi_p(x_T,p_T).
\label{c4}
\end{eqnarray}
Thus the least-action path must be constructed piecewise in each of 
the intervals between observations.

\medskip\noindent
{\bf Example 5.}
To illustrate this situation, we suppose that the positive diffusion
$x$ is a log-Brownian motion:
\begin{equation}
dx_ t = x_t ( \sigma dW_t + \mu dt).
\label{dxgbm}
\end{equation}
Some calculations reveal that the ODE \eqref{c1} is 
\begin{equation}
0 = \frac{p^2 - x\dot{p}+ \sigma^2 x^3}{\sigma^2 x^3}.
\label{ode1}
\end{equation}
Notice that any solution of the ODE \eqref{ode1} must be 
convex while it is positive. 
The non-linear ODE \eqref{ode1} has various solutions, and can indeed be
solved explicitly. However, the explicit solutions are explosive, 
so it turns out that to calculate the solution it  is more effective to 
work numerically, which allows us to truncate the coefficients 
 to prevent explosions. Once a solution is found, the truncation
is no longer necessary.  
Since $\psi_p =(p-\mu x)/\sigma^2x^2 $, the final condition 
\eqref{c4} tells us that we must have $p_T = \mu x_T$, and 
\eqref{c3} gives that
\begin{equation}
   \Delta p_{\tau_i} \equiv p_{\tau_i+} - p_{\tau_i-}
   = - \sigma^2 x_{\tau_i},
   \label{jc}
\end{equation}
which determines the change of derivative over the observation
points. Since the gradient drops at the observation points, this
allows that the solution $x$ found does not have to be convex
globally, even though we have observed that it must be convex in 
each interval between observations.

\begin{figure}[H]
\begin{center}
\includegraphics[width=20cm,height=16cm,angle = 90]{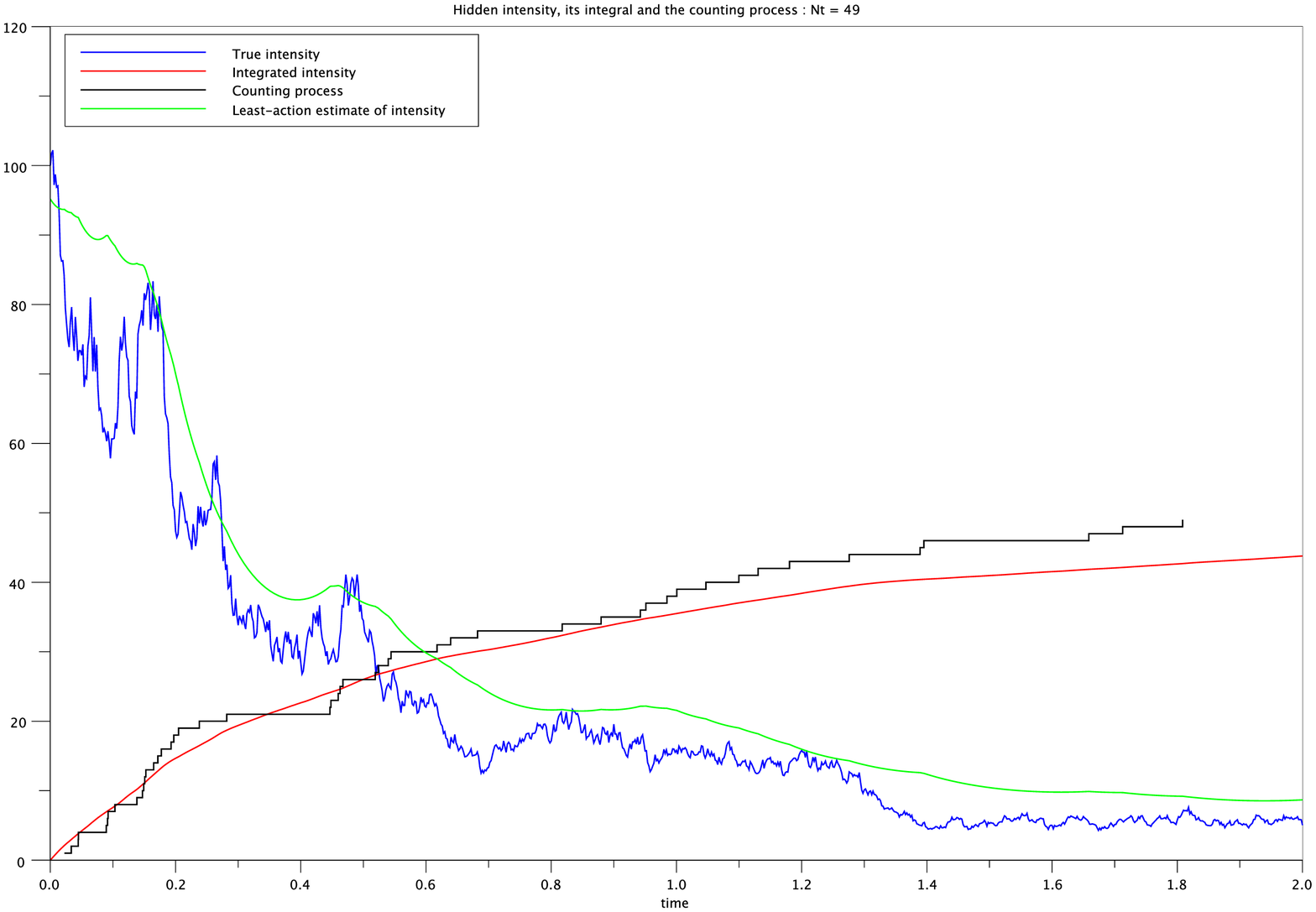}
\caption{An example of least-action estimation of a hidden diffusion.}
\label{pic1}
\end{center}
\end{figure}

 Figure \ref{pic1} illustrates the results of applying the method
 to this example. Despite the fact that there were only 49 points
 observed in the observation period, the general shape of the 
 hidden intensity is well captured.


\section{Relationship to particle filtering.}\label{S5}
For the kind of problems discussed in this paper, the particle
filtering\footnote{The signal processing literature refers to 
{\em Sequential Monte Carlo (SMC)} methods.} 
methodology provides a natural approach. But some simple
thought experiments highlight significant limitations of the 
methodology, and suggests ways in which the approach of this 
paper could in some circumstances supplement or replace 
conventional particle filtering.

Arguably one of the greatest strengths of particle filtering is that
it evolves an estimate of the  {\em  posterior distribution} of the
hidden variables; this guards most effectively against the common
mistake of underestimating the error in parameter estimation.  However,
the methodology though simple is hard to apply effectively.  One place
where problems arise is in evolving the posterior distribution (represented
by a finite collection of $N$ point masses) forward to the next time
point. The simplest version of particle filtering allows each point to 
make a jump  according to the Markovian transition mechanism, and 
then reweights the new particles according to their likelihood given the
new observation.  One often finds\footnote{This is particularly problematic
when there is Gaussian observation error.} that the likelihood of almost
all (or indeed, {\em all}) the new particles is tiny; most of the
new randomly-selected particles are miles away from the observation.

\begin{figure}[H]
\begin{center}
\includegraphics[width=15cm,height=16cm,angle =- 90]{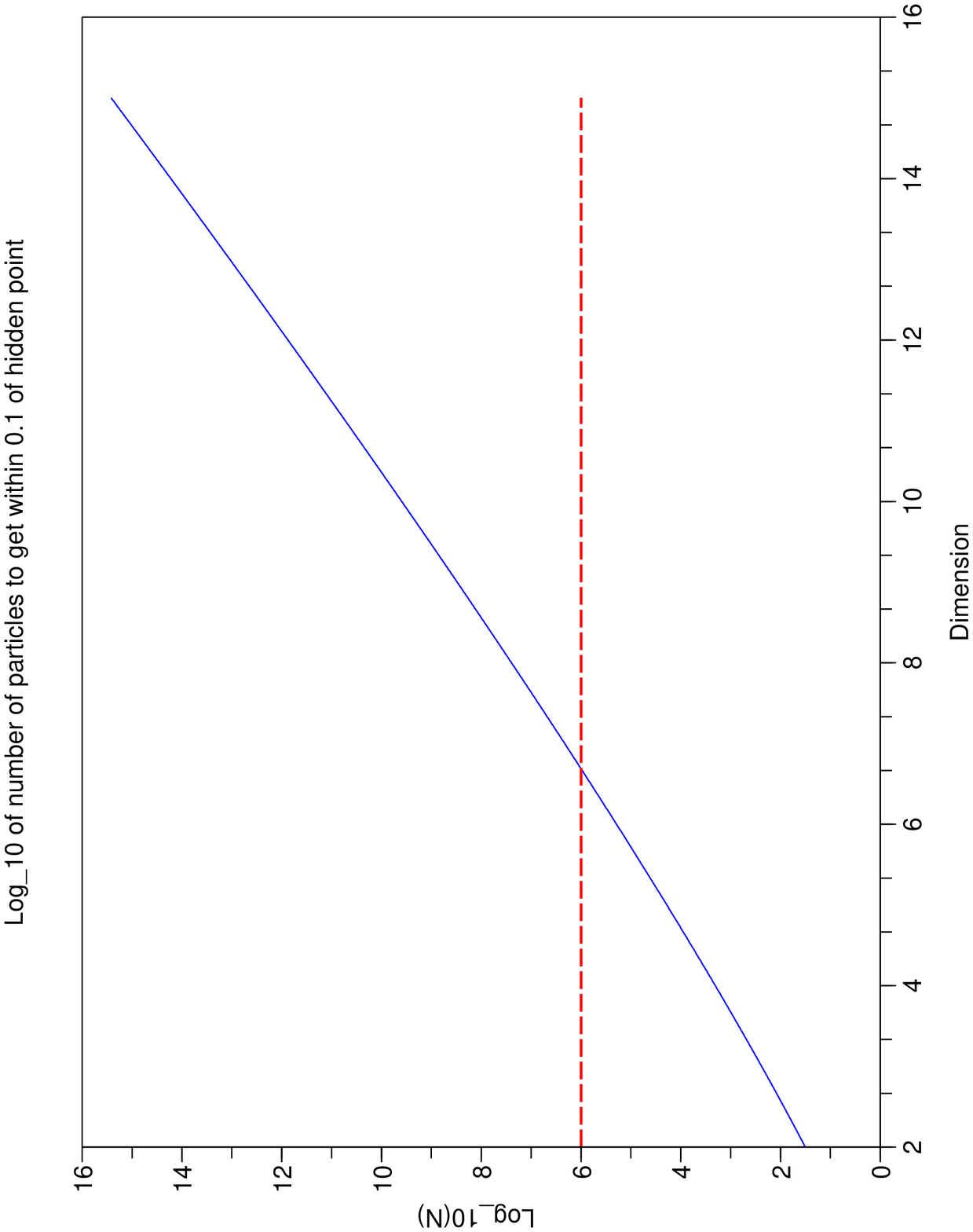}
\caption{Limitations of particle filtering.}
\label{balls}
\end{center}
\end{figure}

To understand the magnitude of the effect, imagine that there is a 
ball $B$ of radius $b$ hidden in the unit cube $C =[0,1]^d$ in $\R^d$; 
think of this ball as the set where $f(\jy|\jx)>\varepsilon$, where 
$f(\cdot|\jx)$ is the density of the new observation given the true
state $\jx$ of the hidden Markov process, and $\varepsilon >0$ is 
some fixed threshold value of likelihood. If points are cast at random
into $C$, how many on average will be needed before one hits the 
ball $B$?  The answer of course is just $b^{-d}V_d^{-1}$, where
\[
    V_d \equiv \frac{(2\pi)^{d/2} \;2^{-d/2} }{ \Gamma(1+\half d ) }
\]
is the volume of the unit ball in $\R^d$.  If we suppose that the 
radius is $b = 0.1$, representing a moderate amount of observational
noise, Figure \ref{balls} plots the mean number $b^{-d}V_d^{-1}$
against dimension $d$, with the line where the mean number is 
one million plotted as a red dashed line across the figure.  This
crosses the plot at dimension roughly equal to 7. Thus if we want 
on average {\em one} particle to fall in the ball of radius 0.1 around the
observation in dimension 7, we will require about one million attempts.
Now of course we would not in practice just mindlessly fire particles
completely at random into $C$; various forms of importance sampling
will hugely improve the success rate.  But this is not quite as 
straightforward as it first appears. If we had a representation for the 
observation as 
\begin{equation}
\jY_t = \jX_t + \jve_t
\label{Yobs}
\end{equation}
where $\jX$ is the hidden state of the Markov process, and $\varepsilon_t$
is a Gaussian observation error, then we would bias the jumps from the 
particles in the population at time $t-1$ towards the new observation 
$\jY_t$.  However, the observation equation \eqref{Yobs} is not how it
is in most examples; more generally we have
\begin{equation}
\jY_t = \Phi(\jX_t) + \jve_t
\label{Yobs2}
\end{equation}
for some very complicated function $\Phi$, and in order to bias the 
jumps of the particles towards values of $\jx$ which are likely relative
to $\jY_t$ {\em we have to know what values of $\jx$ make $\Phi(\jx)$
close to $\jY_t$}.  In other words, we have in effect to find the ML
estimator of $\jX_t$ given $\jY_t$.  Thus the particle-filtering methodology, 
which is wholeheartedly Bayesian in philosophy and execution, is often
forced to resort to frequentist methodology in practice to help it overcome
the curse of dimension.  The least-action approach advanced here is
such a ML method, and could be used to pick out a most likely 
path $\jx^*$ given a  path $\jY$ of observations; the second-order
analysis would tell us how to simulate paths around the most likely
path $\jx^*$ so as to generate a sample of paths which would have
a reasonably high likelihood.

\section{Conclusions.}\label{conc}
This paper offers an approach to estimating a hidden diffusion 
process observed either through other components of a joint
diffusion, or through a point process whose intensity is the hidden
diffusion.   The approach is in effect a maximum-likelihood approach,
but because of the continuous time parameter, calculus-of-variations
techniques can be applied to identify the least-action (=maximum-likelihood)
path.  Moreover, by taking the calculus-of-variations analysis to second
order, we are able to find not only what the most likely path is given
the observations, but also approximately what is the distribution of the 
hidden path about the most likely path.  One point which is worth
emphasizing is that in contrast to other approaches, this methodology
works without any assumption of symmetrizability of the diffusion, and 
requires only calculus in $\R^d$.  The numerical methods involve
shooting solutions of ODEs, so this can be technically quite delicate, 
especially if the time interval over which the solutions are to be constructed
is long.  Accordingly, it cannot yet be claimed that this methodology will
work well in all conceivable situations. However, it is a methodology
which will in principle work in quite high dimension, and for this reason 
shows considerable promise.

\pagebreak
\appendix
\section{Appendix}\label{app}

\medskip\noindent
{\sc Proof of Theorem \ref{thm1}.}
If $\jz : [0,T] \rightarrow \R^d$ is $C^1$, then let $\jz^{(n)}$
denote the piecewise-linear approximation which matches
$\jz$ at each multiple of $h = 2^{-n}$. Then\footnote{Recall
that $s_n = 2^{-n} [2^ns]$.} we have
\begin{eqnarray*}
\lambda_n(\jx^{(n)}|\jy) &=& 
  - \half \sum_{j=0}^{N-1}h \;
    \bigl\vert \; \sigma(jh,\jz^{(n)}_{jh})^{-1}
     \bigl( \; \frac{\jz^{(n)}_{jh+h} - \jz^{(n)}_{jh}}{h}
    - \jmu(jh,\jz^{(n)}_{jh}) \;\bigr) \; \bigr\vert^2 - \varphi(\jx_0)
    \\
    &=&  -\half  \int_0^T  \bigl\vert \; \sigma(s_n,\jz_{s_n})^{-1}
     \bigl( \; \dot{\jz}_{s_n}
    - \jmu(s_n,\jz_{s_n}) \;\bigr) \; \bigr\vert^2 \; ds -\varphi(\jx_0)
    \\
    &\rightarrow & -\half  \int_0^T  \bigl\vert \; \sigma(s,\jz_{s})^{-1}
     \bigl( \; \dot{\jz}_s
    - \jmu(s,\jz_s) \;\bigr) \; \bigr\vert^2 \; ds -\varphi(\jx_0)
    \\
    &=& \Lambda(\jx|\jy)
\end{eqnarray*}
by dominated convergence.  Hence we have
\begin{equation}
    \limsup_{n\rightarrow\infty}  \; \inf_\jx\bigl\lbrace\;
   - \lambda_n(\jx|\jy)\; \bigr\rbrace
  \leq \inf_\jx\bigl\lbrace\; -\Lambda(\jx|\jy)
  \;\bigr\rbrace.
  \label{oneway}
\end{equation}

\noindent
In the other direction, let  $L_n =  \inf_\jx\bigl\lbrace\;
   - \lambda_n(\jx|\jy)\; \bigr\rbrace$, and suppose that $\jx^{(n)}$ is 
   close to minimizing $-\lambda_n(\cdot|\jy)$ in the sense that
\begin{equation}
   -\lambda_n(\jx^{(n)}|\jy) \leq L_n + 2^{-n}.
   \label{eqA1}
\end{equation}   
We then extend $\jx^{(n)}$ by piecewise-linear interpolation
to create a path $\jz^{(n)}$ defined throughout $[0,T]$. We shall
compare the discrete integral
\begin{eqnarray}
D &\equiv &  \half \sum_{j=0}^{N-1}h \;
    \bigl\vert \; \sigma(jh,\jz^{(n)}_{jh})^{-1}
     \bigl( \; \frac{\jz^{(n)}_{jh+h} - \jz^{(n)}_{jh}}{h}
    - \jmu(jh,\jz^{(n)}_{jh}) \;\bigr) \; \bigr\vert^2 
    \nonumber
    \\
    &=& \half  \int_0^T  \bigl\vert \; \sigma(s_n,\jz^{(n)}_{s_n})^{-1}
     \bigl( \; \dot{\jz}^{(n)}_{s_n}
    - \jmu(s_n,\jz^{(n)}_{s_n}) \;\bigr) \; \bigr\vert^2 \; ds
    \label{Ddef}
    \\
    &\equiv & \int_0^T |\ja_s|^2 \; ds
    \nonumber
\end{eqnarray}
with the continuous integral
\begin{eqnarray}
C &\equiv &  \half  \int_0^T  \bigl\vert \; \sigma(s,\jz^{(n)}_{s})^{-1}
     \bigl( \; \dot{\jz}^{(n)}_{s}
    - \jmu(s,\jz^{(n)}_{s}) \;\bigr) \; \bigr\vert^2 \; ds
    \label{Cdef}
    \\
    &\equiv & \int_0^T |\jb_s|^2 \; ds
    \nonumber
\end{eqnarray}
and show that the two are very close for large $n$.  Notice that
the only difference between \eqref{Ddef} and \eqref{Cdef} is that
the time index  passes through the discrete values $j2^{-n}$ in 
the first, but moves continuously through $[0,T]$ in the second. 
The comparison goes via
\begin{eqnarray}
|D-C| &\leq &  \int_0^T | \, |\ja_s|^2 - |\jb_s|^2\, | \; ds
\nonumber
\\
&=&  \int_0^T | (\ja_s - \jb_s) \cdot (\ja_s + \jb_s) | \; ds
\nonumber
\\
&\leq &\biggl( \int_0^T |\ja_s-\jb_s|^2 \; ds  \biggr)
\biggl(  \int_0^T |\ja_s+\jb_s|^2 \; ds   \biggr)
\label{C1}
\end{eqnarray}
by showing that the first factor in \eqref{C1} is small, and the second is
bounded.  Firstly we exploit \eqref{eqA1} to give for some positive finite
constants $\Gamma$  and $\gamma$ and $0 \leq t \leq u \leq T$  the bounds
\begin{eqnarray*}
\Gamma &\geq & \int_t^u  \bigl\vert \; \sigma(s_n,\jz^{(n)}_{s_n})^{-1}
     \bigl( \; \dot{\jz}^{(n)}_{s_n}
    - \jmu(s_n,\jz^{(n)}_{s_n}) \;\bigr) \; \bigr\vert^2 \; ds
    \\
    &\geq & \gamma  \int_t^u  \bigl\vert \;
      \dot{\jz}^{(n)}_{s_n}
    - \jmu(s_n,\jz^{(n)}_{s_n})  \; \bigr\vert^2 \; ds
    \\
    &\geq & \gamma  \int_t^u   \biggl\lbrace\,  \half  \bigl\vert \;
  \dot{\jz}^{(n)}_{s_n} \bigr\vert^2 
    - \bigl\vert \; \jmu(s_n,\jz^{(n)}_{s_n})  \; \bigr\vert^2
    \, \biggr\rbrace  \; ds.
\end{eqnarray*}
Hence we have the inequality\footnote{$\Gamma$ denotes as
usual a positive finite constant whose value matters little, and 
may change from line to line.}
\begin{equation}
 \int_t^u \bigl\vert \; \dot{\jz}^{(n)}_{s_n}  \; \bigr\vert^2 \; ds
 \leq \Gamma
 \label{ineqA2}
\end{equation}
which yields the modulus of continuity
\begin{eqnarray}
 |\jz^{(n)}_u - \jz^{(n)}_t|^2 &=&\bigl\vert \;  \int_t^u
              \dot{\jz}^{(n)}_{s_n}  \;  \; ds\bigr\vert^2
              \nonumber
              \\
              &\leq & \Gamma\,  \sqrt{u-t}.
              \label{ineqA3}
\end{eqnarray}
It is easy to see that the analogous inequality 
\begin{equation}
 \bigl\vert \;  \int_t^u
              \dot{\jz}^{(n)}_{s}  \;  \; ds\bigr\vert^2
             \leq \Gamma\,  \sqrt{u-t}.
             \label{ineqA4}
\end{equation}
also holds.  The boundedness of the second factor in \eqref{C1}
now follows. Using the modulus of continuity, it is now straightforward
to show that the first factor in \eqref{C1} can be made arbitrarily
small by taking $n$ arbitrarily large.

Hence given $\varepsilon>0$,  for large enough $n$ we have 
\begin{equation*}
\inf_\jx\bigl\lbrace\; -\Lambda(\jx|\jy)
  \;\bigr\rbrace \leq  -\Lambda(\jx^{(n)} |\jy)
  \leq  -\lambda_n(\jx^{(n)}|\jy) + \varepsilon
   \leq L_n + 2^{-n}+\varepsilon.
\end{equation*}
We conclude that
\begin{equation*}
   \inf_\jx\bigl\lbrace\; -\Lambda(\jx|\jy)\;\bigr\rbrace
    \leq \liminf_{n\rightarrow\infty}
   L_n
\end{equation*}
which combines with \eqref{oneway} to give the statement \eqref{t1}.

For the final statement of Theorem \ref{thm1}, we use the condition
\eqref{regcon} along with the modulus of continuity proved above. This
shows that the family $\{ \jx_n\}$ is equicontinuous, so by the Arzela-Ascoli
Theorem is relatively compact in the topology of $C[0,T]$. Any subsequential
limit of the $\jx_n$ converges to a path  for which the value of 
$-\Lambda(\jx|\jy)$ is minimal; but by assumption there is only one
such path, $\jx^*$, and so any subsequential limit of the $\jx_n$ 
must be $\jx^*$, hence the $\jx_n$ converge in the (uniform) topology
of $C[0,T]$ to $\jx^*$.

\hfill $\square$

\bibliography{LAFbib}

\end{document}